# Hetero-Orbital Two-Component Fractional Quantum Hall States in Bilayer Graphene

Ke Huang[1], Ajit C. Balram[2,3], Hailong Fu[1†], Chengqi Guo[1], Kenji Watanabe[4], Takashi Taniguchi[5], Jainendra K. Jain[1*], Jun Zhu[1,6*]

**Affiliations**

[1]Department of Physics, The Pennsylvania State University, University Park, Pennsylvania 16802, USA.
[2]Institute of Mathematical Sciences, CIT Campus, Chennai 600113, India.
[3]Homi Bhabha National Institute, Training School Complex, Anushaktinagar, Mumbai 400094, India
[4]Research Center for Electronic and Optical Materials, National Institute for Materials Science, 1-1 Namiki, Tsukuba 305-0044, Japan
[5]Research Center for Materials Nanoarchitectonics, National Institute for Materials Science, 1-1 Namiki, Tsukuba 305-0044, Japan
[6]Center for 2-Dimensional and Layered Materials, The Pennsylvania State University, University Park, Pennsylvania 16802, USA.
[*]Correspondence to: jxz26@psu.edu (J. Zhu) and jkj2@psu.edu (J. K. Jain)
[†] Present address: School of Physics, Zhejiang University, Hangzhou 310058, China.

**Abstract**

A two-dimensional electron system exposed to a strong magnetic field produces a plethora of strongly interacting fractional quantum Hall (FQH) states, the complex topological orders of which are revealed through exotic emergent particles, such as composite fermions, and fractionally charged Abelian and non-Abelian anyons. Much insight has been gained by the study of multi-component FQH states, where spin and pseudospin indices of the electron contribute additional correlation. Traditional multi-component FQH states develop in situations where the components share the same orbital states and the resulting interactions are pseudospin independent; this homo-orbital nature is also crucial to their theoretical understanding. Here, we study "hetero-orbital" two-component FQH states, in which the orbital index is part of the pseudospin, rendering the multi-component interactions strongly SU(2) anisotropic in the pseudospin space. Such states, obtained in bilayer graphene at the isospin transition between N = 0 and N = 1 electron Landau levels, are markedly different from previous homo-orbital two-component FQH states. In particular, we observe strikingly different behaviors for the parallel-vortex and reverse-vortex attachment composite fermion states, and an anomalously strong two-component 2/5 state over a wide range of magnetic field before it abruptly disappears at a high field. Our findings, combined with detailed theoretical calculations, reveal the surprising robustness of the hetero-orbital FQH effects, significantly enriching our understanding of FQH physics in this novel regime.

## I. Introduction

Landau Levels (LLs) of a two-dimensional electron gas support a plethora of strongly correlated electronic phases, where the form factor of the orbital wave function plays a crucial role in determining the characteristics of the underpinning Coulomb interactions [1,2]. In GaAs quantum wells, FQH states of the sequence $\nu$ = p/(2p ± 1) occur in the n = 0 LL and are well described by the integer QH states of composite fermions (CFs), namely particles produced by binding of electrons and two vortices, consisting of p-filled CF Landau-levels called Λ levels [1]. However, the origin of the FQHE in the n = 1 LL is very different. Here a softened Coulomb interaction leads to the pairing of CFs at half-fillings to produce putatively non-Abelian FQH states, such as the 5/2 state in GaAs [3]. Unconventional FQH states at partial fillings $\tilde{\nu}$ = 2/5 and 3/8 in the n = 1 LL have also been hypothesized to be non-Abelian [3-12]. Exploring the interplay of the n = 0 and n = 1 LLs through level crossing offers valuable insights into the physics of the FQHE [13-19]. For example, it was reported that the 5/2 state in a wide quantum well strengthens briefly before crossing to a n = 0 compressible liquid [13,18].

The study of multi-component two-dimensional (2D) systems, where the components – generically referred to as pseudospin – may be the electron spin [1,2,19-27], quantum well/atomic layer [28-32], valley/sublattice isospin [33-39] or subband [18,40,41], further enriches the FQHE by establishing different correlations between different components. Double-layer systems exhibit superfluid behavior believed to be described by the Halperin (111) wave function [28,30,31,42], where both intralayer and interlayer interactions lead to the formation of a collective $\nu$ = 1 state. The model of spinful CFs captures the spin/valley singlet, e.g. the equivalent of the Halperin (332) state at ν = 2/5, or partially spin polarized states in the Jain FQH sequence of $\nu$ = p/(2p ± 1) [1,2,19-27,33,36,39]. It is interesting to note that all established two-component states so far are homo-orbital in nature, i.e. have the same LL wave function in both components. Hetero-orbital two-component states, while not theoretically prohibited, have not been observed.

The eight closely degenerate LLs of bilayer graphene centered at the charge neutrality point ($E$ = 0) offer a rich platform to advance the physics of multi-component FQHE thanks to its inherent spin/valley/orbital indices and wide tunability [34-37,39]. While monolayer graphene exhibits the conventional Jain states at $\nu$ = p/(2p ± 1) in both N = 0 and 1 LLs [21,22], BLG exhibits the Jain states in the N = 0 LL and even-denominator FQH states in the N = 1 LL [35,36,39]. FQH states at 2/5 and 3/7 have also been observed in the N = 1 LL, although their nature remains an open question [35,36,39]. An electric displacement field ($D$-field) controls the valley Zeeman splitting, which in turn induces the crossing of LLs carrying different orbital (N = 0,1) and valley (ξ = +, −) indices as illustrated in Fig. 1(a) [39]. This crossing point is an excellent place to explore multi-component states. The orbital wave function of the N = 1 LL in BLG comprises a majority n = 1 orbital and a minority n = 0 orbital in GaAs (Fig. 1(a)), with the weight of the latter increasing smoothly with increasing magnetic field; this unique property allows for the exploration of the orbital wave-function-driven topological phase transitions, as predicted recently [12,16,36].

In this work, we report on the observations of hetero-orbital two-component FQH states occurring at the isospin transition of the $|\pm 0\rangle$ and $|\mp 1\rangle$ electron LLs in bilayer graphene. Their behavior is underpinned by anisotropic SU(2) interactions and exhibits properties remarkably different from those seen in past homo-orbital systems. A single hetero-orbital two-component state occurs for two-vortex CF states at partial fillings 2/5 and 3/7, but not for reverse vortex CF states at p/(2p – 1). This is in stark contrast to homo-orbital two-component systems where FQH states are observed at both p/(2p + 1) and p/(2p – 1) fillings [1,2,20,33,36,39]. A systematic study of the magnetic-field-dependent energy gaps of the three FQH phases at filling factor 7/5 reveals a number of unusual observations. The very strong hetero-orbital two-component 2/5 state, denoted as the $D^*$ state in measurements, develops at a low field of $B \sim 7$ T, and has the largest gap $\Delta_{2/5}^{D^*} > \Delta_{2/5}^{N=0} > \Delta_{2/5}^{N=1}$, but this state disappears at ~ 31 T. The gap of the conventional FQH phase on N = 0 LL follows the $\sqrt{B}$ scaling and is characterized by an effective CF mass of $0.13m_e$. The gap of the N = 1 phase $\Delta_{2/5}^{N=1}$ increases rapidly with $B$ at lower field but slows down to merge with $\Delta_{2/5}^{N=0}$ at around 28 T, where the two-component state begins to disappear. Exact diagonalization calculations validate the stabilization of an SU(2) anisotropic two-component state at p/(2p + 1) at the $|\pm 0\rangle/|\mp 1\rangle$ coincidence and capture the absence of such a state at p/(2p – 1), but they do not explain the collapse of the two-component state at a high $B$. Our work opens a new avenue to explore multi-component FQH states in bilayer graphene, a high-quality 2D electron gas with a rich isospin structure and immense tunability.

## II. Experimental results

Our measurements employ high-quality graphite dual-gated, h-BN encapsulated Hall bar devices, the fabrication and characterization of which are described in our previous work [39] and in Appendix A. The small $D$-field inhomogeneity ($\delta D$ < 0.7 mV/nm) realized in our device is key to the observation in this work. Figure 1(b) shows a false color map of the longitudinal resistance $R_{xx}(D, \nu)$ near $\nu$ = 3/2 in device 002, where the FQH states appear as vertical dark lines. Many observations of this map were discussed in Ref.[39]. Here we focus on features near the crossing of the $|+0\rangle$ and $|-1\rangle$ LLs which occurs at a set of ν-dependent $D$-field values, $D^*$, as illustrated by the dashed line in the schematic diagram below. In its vicinity, fractional states of filling factor range $1 < \nu < 2$ occupy the $|+0\rangle$ LL at low $D$ and transition to the $|-1\rangle$ LL at high $D$ in two different ways. For states with $\nu$ > 3/2, e.g. 8/5 and 11/7, $R_{xx}$ exhibits a resistance peak at $D^*$ (see Fig. 1(c) and Fig. 5, Appendix C). This peak is commonly observed at LL crossings and corresponds to a brief closing of the gap. In stark contrast, at $\nu$ = 7/5, 10/7, and 13/9, a new incompressible state with a deep $R_{xx}$ minimum emerges at $D^*$. In Fig. 1(d), we verify that the $D^*$ state of 7/5 is a FQH state by observing a Hall conductance plateau at the correct value. A shoulder also develops for the $D^*$ state of 10/7. Figure 1(c) shows the evolution of $R_{xx}$ as a function of $D$ at fixed $\nu$ = 7/5, 10/7, 13/9, 11/7, and 8/5 respectively. Traces at $\nu$ = 7/5, 10/7, and 13/9 exhibit a clear $R_{xx}$ minimum at its own $D^*$. The $D^*$ state is separated from the FQH states occupying the $|+0\rangle$ and $|-1\rangle$ LLs by two resistance peaks. The $|+0\rangle$ states belong to the conventional Jain FQH sequence while the nature of the $|-1\rangle$ states (at $\nu$ = 7/5, 10/7) remains to be clarified [39]. We observe the $D^*$ state clearly and consistently at 7/5 and 10/7 in multiple devices and in a wide range of $B$-fields. Additional data are given in Fig. 5, Appendix C. Its appearance supports the development of a new incompressible FQH state at the $|+0\rangle/|-1\rangle$ coincidence, and similarly at $|-0\rangle/|+1\rangle$ on the negative $D$ side. However, we have not seen this state at fractional fillings $\nu$ > 3/2, such as 11/7 and 8/5. Clearly, CF states of parallel-vortex ($\nu$ = p/(2p + 1)) and reverse vortex ($\nu$ = p/(2p − 1)) attachment behave very differently here, in contrast to past spin/valleyful CF systems [1,2,20,33,36,39].

We examine the evolution of the $D^*$ state in a magnetic field, using $\nu$ = 7/5 in device 002 as an example. Figure 2(a) shows a series of $D$-sweeps at $\nu$ = 7/5 and representative $B$-fields varying from 7 to 31 T. A full $B$-field set is given in Fig. 6, Appendix C. It is quite remarkable that the $D^*$ state appears *stronger* than the conventional FQH phase riding on the N = 0 LL and is already well developed at 7 T. Interestingly, $\Delta D$, the $D$-field range occupied by the $D^*$ state, which is marked by the distance between the flanking $R_{xx}$ peaks in Fig. 2(a), exhibits a non-monotonic magnetic field dependence, as shown in Fig. 2(b). In device 002, the $D^*$ state forms between approximately 6 and 30.5 T with a maximum range $\Delta D$ = 2 mV/nm at $B$ = 16 T. In Fig. 2(c), we plot an experimental $D$-$B$ phase diagram for $\nu$ = 7/5, where it exhibits three FQH phases with distinct orbital wave functions. The disappearance of the $D^*$ state at $B$ > 30 T, which also occurs for the $D^*$ state at $\nu$ = 10/7 and 13/9 and in device 011 (Fig. 7, Appendix C), is unusual, as FQH states are typically strengthened at large $B$ due to stronger Coulomb interactions. We note that the disappearance is unlikely caused by a spin polarization transition. The significant exchange-enhanced spin Zeeman splitting compared to the small valley splitting in the large $B$ and small $D$ regime studied here makes the spin degrees of freedom non-active in this problem [35,37,39]. These unprecedented observations and the intricate $D$ and $B$ dependences attest to a rich and tunable interaction landscape in BLG, where new FQH states emerge.

To further understand the three FQH phases of 7/5, we have systematically measured their energy gaps in a wide range of magnetic field using the temperature dependence of $R_{xx}$, an example of which is given in Fig. 3(a) for $B$ = 22 T. It is immediately clear from the data that the $D^*$ state is the strongest, while the N = 1 state is the weakest. Figure 3(b) shows a few exemplary Arrhenius plots $R_{xx} \sim \exp(-\Delta/2k_BT)$, from which we determine the gap size $\Delta$. Values of $\Delta_{2/5}^{D^*}$, $\Delta_{2/5}^{N=0}$, and $\Delta_{2/5}^{N=1}$ are extracted and plotted as a function of $B$ in Fig. 3(c) from 14 to 31 T while additional measurements and fits are given in Figs. 8-10, Appendix C. In our previous work, a non-interacting two-component CF model describes the valley isospin polarization transitions near $D$ = 0 very well [39]. Thus, we have adopted a similar model here to fit the gap of the N = 0 state to $\Delta_{2/5}^{N=0} = \hbar e B_{\text{eff}}/m_a^{\text{CF}} - \Gamma$, where $m_a^{\text{CF}} = \alpha m_e \sqrt{B}$ ($m_e$ is the free electron mass) and $\Gamma$ is the disorder broadening . From the fitting, we extract $\alpha$ = 0.13 and $\Gamma$ = 6.8 K (See analysis in Fig. 10, Appendix C) [1,22,43]. The $m_a^{\text{CF}}$ value is larger than that ($m_a^{\text{CF}} = 0.067 m_e \sqrt{B}$) obtained for the 2/5 state in monolayer graphene [22], suggesting that LL mixing plays a more important role in BLG; a similar conclusion is reached in [39]. While we do not have a quantitative theory, it is truly remarkable that the gap of the $D^*$ phase exceeds that of the conventional N = 0 phase by more than one Kelvin. This finding points to strong correlations at the $|\pm 0\rangle/|\mp 1\rangle$ coincidence.

The N = 1 phase of 7/5 exhibits a more complex field evolution. Generally, conventional FQH states are not stabilized on the n = 1 LL of GaAs [4,5]. However, the N = 1 LL in BLG consists of a small n = 0 component, the weight of which increases with increasing $B$ (Fig. 1(a)). Experimentally, the N = 1 phase of 7/5 only develops at a substantial $B$-field. Once formed, $\Delta_{2/5}^{N=1}$ rapidly increases with $B$, reaches the size of $\Delta_{2/5}^{N=0}$ at around 28 T, and then merges with the trace of $\Delta_{2/5}^{N=0}(B)$ (Fig. 2(c)). This $B$-dependence is consistent with previous capacitance measurements [36] and may be due to the unique $B$-field evolution of the N = 1 wave function in BLG. Whether the N = 1 phase of 7/5 is Abelian or non-Abelian remains an open question. It is interesting that both the coalescence of $\Delta_{2/5}^{N=1}$ and $\Delta_{2/5}^{N=0}$ and the disappearance of the $D^*$ state occur at around 28 T, suggesting a substantial change of the underpinning interactions that are important to both phenomena.

### III. Theoretical calculations and discussions

What is the nature of the new $D^*$ state? Given its occurrence at the $|\pm 0\rangle/|\mp 1\rangle$ level crossing, a two-component state seems a natural candidate. In an earlier work [39], we showed that a two-component CF model composed of a $(|+0\rangle, |-0\rangle)$ spinor can quantitatively describe the many valley isospin polarization transitions that occur near $D = 0$ [1,2,24-27,44]. There, a state of $\nu = p/(2p \pm 1)$ maps to p-filled CF Λ levels and exhibits p level crossings and p – 1 two-component states from $-D$ to $+D$. A direct generalization of this scheme to a $(|\mp 1\rangle,\ |\pm 0\rangle)$ spinor would imply one $D^*$ state for $\nu$ = 2/5 and 2/3, two $D^*$ states for 3/7 and 3/5 and so on. Experimentally, however, we only observed a single $D^*$ state at $\nu$ = 2/5, 3/7, and 4/9, and none at the reverse-vortex states. This qualitative discrepancy points to the deficiency of a conventional two-component model. In past homo-orbital two-component systems, the diagonal SU(2) interaction matrix elements are identical, i. e. $V^{\uparrow,\uparrow}(r) = V^{\downarrow,\downarrow}(r)$ [1,2,24-27,42]. Here, the hetero-orbital nature of our system leads to anisotropic interaction matrix elements: $V^{+0,+0}(r) \neq V^{+0,-1}(r) = V^{-1,+0}(r) \neq V^{-1,-1}(r)$, thus the stability of two-component FQH states must be reexamined. It is worth noting that incompressible states have also been observed at the crossing of the N = 0 and 1 LLs near $\nu = -5/2$ in BLG, although their overall characteristics seem to follow the conventional homo-orbital two-component model quite well [36].

We quantitatively investigated this problem by obtaining the exact ground states of the anisotropic two-component system through exact diagonalization and comparing the solutions to the two-component spin-singlet $\Psi_{2/5}^{(1,1)}$, $\Psi_{2/3}^{(-1,-1)}$, and partially spin-polarized $\Psi_{3/7}^{(2,1)}$ and $\Psi_{3/5}^{(-2,-1)}$ Jain states that provide near-exact representations for the SU(2) symmetric Coulomb interaction. The interactions used are given in Appendix B. Note that our model does not include LL mixing, lattice-scale anisotropies or hopping terms beyond the nearest neighbor (that result in trigonal warping). Figure 4(a) plots the wave function overlaps for $\nu$ = 2/5 and 2/3, which correspond to 2 and −2 filled CF LLs (the latter with reverse vortex attachment). The overlaps for $\nu$ = 3/7 and 3/5, corresponding to CF fillings of 3 and −3, are given in Fig. 11, Appendix C. We find that for a broad range of $B$-field, the exact two-component ground states of the anisotropic Hamiltonian at $\nu$ = 2/5 and 3/7 are almost perfectly described by the *isotropic* CF states, implying that the ground state is SU(2) symmetric to an excellent approximation even though the microscopic interaction is not. In contrast, the agreement is relatively poor for $\nu$ = 2/3 and $\nu$ = 3/5, indicating that two-component FQHE is not robust and may not be realized experimentally at these filling factors. These findings agree remarkably well with the experiment. In the case of $\nu$ = 2/5, our results suggest that the $V_0$ and $V_1$ Haldane pseudopotentials are positive and dominant for each of the $V^{\sigma,\sigma'}$ interactions ($\sigma, \sigma'$ = +0, −1) since $\Psi_{2/5}^{(1,1)}$ is the exact ground state of a model with only $V_0$ and $V_1$ terms. We do not have similar qualitative arguments for why the state at 3/7 is well described by the Jain wave function but those at 3/5 and 4/7 are clearly not.

Further insights into the hetero-orbital two-component states can be gained by comparing the phase diagrams of the isotropic and anisotropic interactions. Figures 4(b) and (c) illustrate the key physics at $\nu$ = 2/5. The calculated ground state energies of the single and two-component states $E(n_\uparrow, n_\downarrow)$, where $n_{\uparrow,\downarrow}$ is the filling factor of the N = $|\mp1\rangle$ and $|\pm0\rangle$ orbitals respectively, evolve linearly with the $D$-field, and their competition determines which state manifests. In the isotropic case, $E(2, 0) = E(0, 2)$ at $\Delta D = D - D_s^* = 0$, where $D_s^*$ is the single-particle LL crossing point, and $E(1, 1)$ is below both E(2,0) and E(0,2) ; all three become ground states as a function of $\Delta D$, as shown in the first diagram of Fig. 4(c). In the anisotropic case, however, the energy ordering of these states is qualitatively different. We have calculated $E(2, 0)$, $E(1, 1)$ and $E(0, 2)$ as a function of the magnetic field by extrapolating the finite system results to the thermodynamic limit and plotted the results in Fig. 12, Appendix C. Our calculations show that $E(2, 0) >$ $E(0, 2)$ and (1, 1) lies in between. If $E(1, 1)$ is closer to $E(0, 2)$ than to $E(2, 0)$, (1, 1) will become the ground state for some range of $\Delta D$, as depicted in Anisotropic I of Fig. 4(c). Our calculations corroborate this scenario and we identify the experimental $D^*$ state with the two-component (1, 1) state. If $E(1, 1)$ is closer to $E(2, 0)$ instead, the system will transition directly from the (0, 2) state to the (2, 0) state without going through the $D^*$ state. We suspect that this situation, called Anisotropic II, corresponds to the experimental situation of $B$ > 30 T though this disappearance is not captured by our calculations. Concomitantly with the disappearance of the $D^*$ state, the slope of $\Delta_{2/5}^{N=1}(B)$ also changes near 28 T. These observations, together with the large value of $\Delta_{2/5}^{D^*}$ and the substantial renormalization of $\Delta_{2/5}^{N=0}$ , point to the need for a more refined theoretical treatment to capture the energetics of BLG quantitatively.

Similar considerations can also be applied to $\nu$ = 3/7 to explain the formation of a single $D^*$ state at the level crossing. In an isotropic two-component system, four ground states including two partially polarized states are expected and indeed observed at $\nu$ = 11/7 = 2 − 3/7 near $D$ = 0 [39]. Here anisotropic interactions lead to the splitting between $E(3, 0)$ and $E(0, 3)$ and between $E(2, 1)$ and $E(1, 2)$. Exact diagonalization calculations performed for the 3/7 state are given in Fig. 11, Appendix C, where we show how the movement of the energies allows the possibility of eliminating the partially polarized (2, 1) state so that only the (1, 2) state survives in the vicinity of the crossing. This is likely the experimental $D^*$state we observed. Its disappearance at high $B$ likely originates from a similar interaction change to the 2/5 case. Indeed, the $D^*$states at 7/5, 10/7, and 13/9 all disappear around 30 T in our measurement (Fig. 7, Appendix C), which we hope that future calculations may be able to capture.

## IV. Summary and outlook

The multiple isospin degeneracy and strong correlations in the $E$ = 0 Landau level of graphene and bilayer graphene offer an excellent platform to explore the physics of many-body coherent, multi-component states, such as a canted anti-ferromagnet and a Kekulé phase at ν = 0 [45-48]. In this work, we combined experiment and theory to reveal the properties of unconventional two-component FQH states, occurring at the coincidence of $|\mp1\rangle$ and $|\pm0\rangle$ Landau levels in bilayer graphene. We show that even if the intra- and inter-component interactions are all different, two-component states can still be stabilized, albeit with a set of characteristics distinguished from previous isotropic systems. Our measurement of the energy gaps of the three FQH phases at $\nu = 7/5$ also highlight the importance of understanding the impact of Landau level mixing in bilayer graphene, in order to reach agreement with experiments. Finally, the complex magnetic field evolution of fractional states riding on the $N = 1$ Landau level, and the possibility of other topological orders [12], remain an open question for future experiment and theory.

**Figures:**

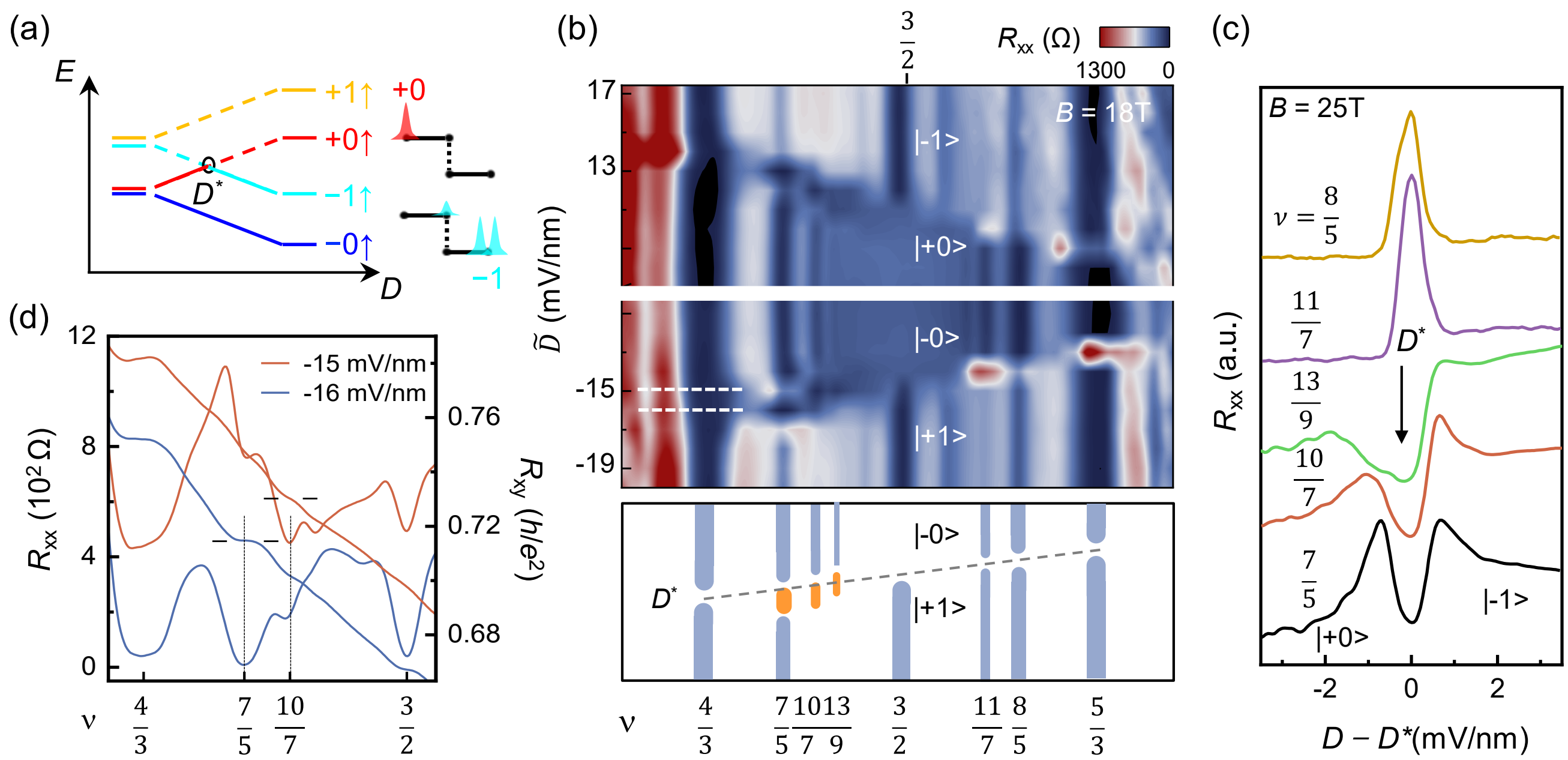

Fig. 1. Two-component FQH states at the coincidence of $|\pm0\rangle$ and $|\mp1\rangle$ LLs (a) A schematic $E(D)$ diagram showing the electron LLs and the $|+0\rangle/|-1\rangle$ coincidence at positive $D^*$ (black circle). Each LL is labeled by $|\xi N\sigma\rangle$, where $\xi = +, -$, $N = 0, 1$, and $\sigma = \uparrow$ and $\downarrow$ denote the valley, orbital and spin indices respectively. The cartoon on the right illustrates the atomic site wave function distribution of the $|+0\rangle$ and $|-1\rangle$ LLs. (b) Upper panel: A false color map of $R_{xx}(\widetilde{D}, \nu)$. $\widetilde{D} = D + 5$ mV/nm adjusts for the offset in the nominal applied $D$-field. FQH states appear as vertical dark lines. Lower panel: A schematic representation of the data taken at $D < 0$. The dashed line marks the $D^*$ in the range $4/3 < \nu < 5/3$. New FQH states, which appear as dark spots in the upper panel and are colored orange in the lower panel, emerge at $D^*$ for $\nu = 7/5$, $10/7$, and $13/9$, but not for $\nu = 4/3$, $11/7$, $8/5$, and $5/3$. (c) Vertical scans of $R_{xx}(D)$ taken along $\nu = 7/5$, $10/7$, and $13/9$ (device 011), and $11/7$ and $8/5$ (device 002) respectively and centered at their respective values of $D^*$. The $D^*$ state occupies a similar $D$-field range at $\nu = 7/5$, $10/7$, and $13/9$. It is absent at $\nu = 8/5$, and $11/7$, similar to what's shown in the map in (b). $B = 25$T, $T = 0.33$ K. See Fig. 5 for additional measurements on the $D^*$ states at different magnetic fields and in different devices. (d) Horizontal scans of $R_{xx}(\nu)$ and $R_{xy}(\nu)$ taken at a fixed $\widetilde{D} = -15$ (red) and $-16$ (blue) mV/nm as indicated by the white dashed lines in (b). The former cuts through the $D^*$ state of 10/7 and 13/9, and the latter cuts through the $D^*$ state of 7/5 on the negative $D$ side. The $D^*$ state of 7/5 exhibits a well-developed plateau at $R_{xy} = 5/7\ h/e^2$. The $D^*$ state of 10/7 shows an incipient plateau at $R_{xy} = 7/10\ h/e^2$. The red $R_{xx}$ and $R_{xy}$ traces are offset by 400 Ω and 0.03 $h/e^2$ respectively for clarity. From device 002. $B = 18$ T and $T = 20$ mK.

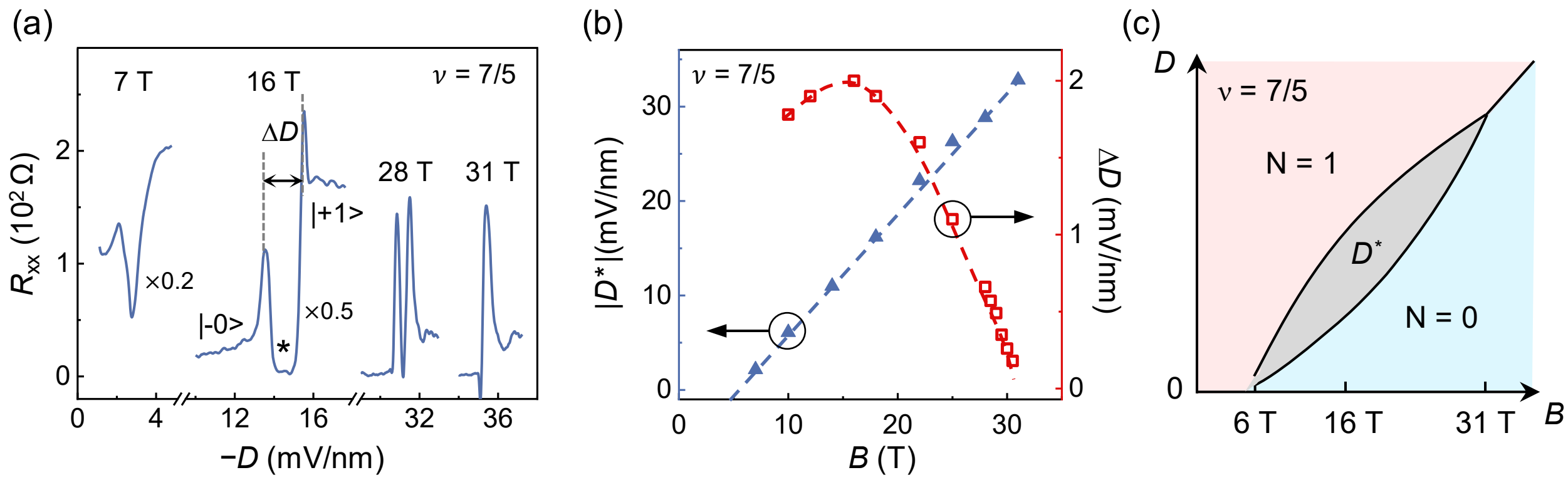

Fig. 2. Magnetic field evolution of the $\nu$ = 7/5 $D^*$ state. (a) $R_{xx}^{7/5}(D)$ sweeps at selected *B*-fields as labeled. Additional *B*-fields are given in Fig. 6. (b) The position of $D^*$ (left axis, blue triangle) and the phase space Δ*D* (right axis, red open square) as a function of *B*. Δ*D* measures from peak to peak, as illustrated in (a). $|D^*|(B)$ follows a linear *B*-dependence given by $|D^*| = 1.3B\text{-}6.8$ (dashed line). Data in (a) and (b) are taken from the negative *D* side of device 002. (c) A schematic *D*-*B* phase diagram of the 7/5 showing the three FQH phases with different orbital wave functions. The phase space of the $D^*$ state is exaggerated for clarity.

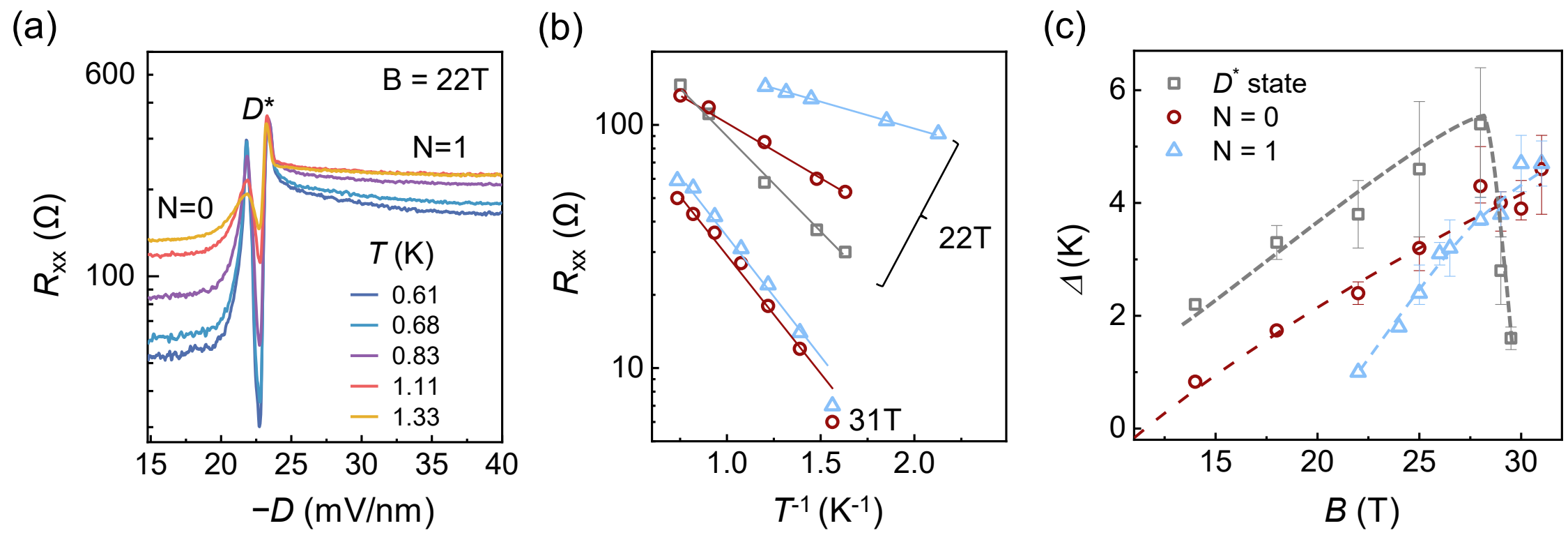

Fig. 3. Magnetic field dependent energy gaps of the three FQH phases at 7/5. (a) $R_{xx}^{7/5}(D)$ sweeps at selected temperatures as labeled in the plot. $B$ = 22 T. $R_{xx}$ becomes $D$-independent when the state is fully riding on N = 0 or 1 LLs. We extract the $T$-dependent $R_{xx}(D^*)$, $R_{xx}$(N = 0), and $R_{xx}$(N = 1) from similar traces and plot $R_{xx}(T)$ in an Arrhenius plot. (b) Exemplary Arrhenius plots for the three states at selected $B$-fields as labeled. Symbols follow the legend in (c). Additional measurements and analyses are given in Figs. 8-10. (c) Magnetic field dependence of $\Delta_{2/5}^{N=1}$ (blue triangle), $\Delta_{2/5}^{N=0}$ (red circle), and $\Delta_{2/5}^{D^*}$ (gray square). The red dashed line is a fit to data: $\Delta_{2/5}^{N=0}(K) = 2.0\sqrt{B(T)} - 6.8$. See Fig. 10 for details. Blue and gray dashed lines are guide to the eye. From device 002.

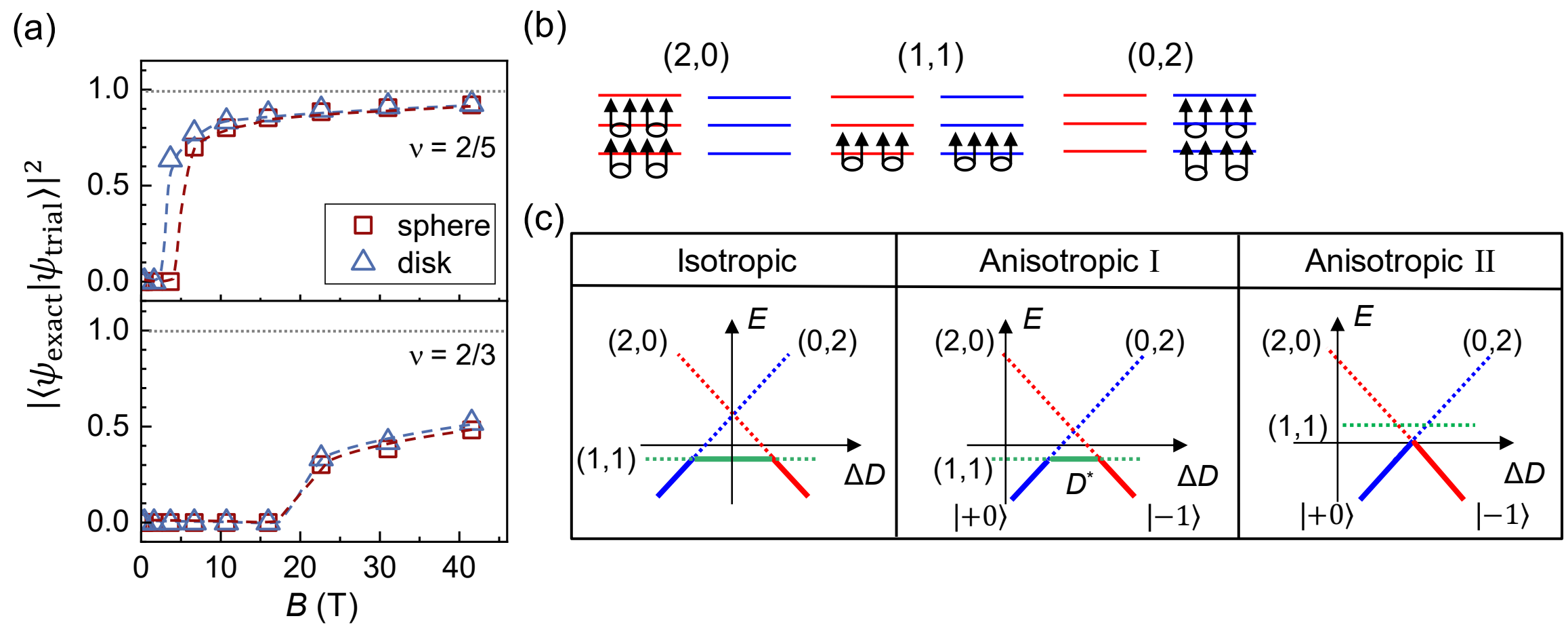

Fig. 4. Theoretical calculations of anisotropic two-component states at ν = 2/5 and 2/3. (a) Overlaps of the spin-singlet Jain state ($|\psi_{trial}\rangle$) with the exact Coulomb ground state ($|\psi_{exact}\rangle$) in the $E$ = 0 LL of BLG as a function of the magnetic field for $\nu$ = 2/5 (upper panel) and 2/3 (lower panel). For the Jain states, we have used the Coulomb ground state that arises for $B$ -> infinity in the N = 1 LL of BLG, which is identical to that in the n = 0 LL. Calculations are performed for n electrons in the spherical geometry using the spherical (red sphere) and disk (blue triangle) pseudopotentials. See Appendix B for the interaction details. The electron number n = 10 for $\nu$ = 2/5 and 14 for = 2/3. (b) CF Λ level filling diagram for the (2, 0) (N = $|-1\rangle$ ), (1, 1) (two-component), and the (0, 2) (N = $|+0\rangle$) states of the 2/5 respectively. (c) Schematic energy diagrams showing the evolution of $E$(2, 0), $E$(1, 1) and $E$(0, 2) in a $D$-field and the resulting ground states. Three scenarios are given. In the first diagram, an isotropic interaction results in three ground states with $D^*$ being the (1, 1) state. In the second diagram, we show how an anisotropic interaction shifts the location of $D^*$ and changes the width of the (1, 1) state. The third diagram depicts a scenario where the (1, 1) state never becomes a ground state. This could potentially explain the disappearance of the $D^*$ state at very large $B$.

**Acknowledgement**

We thank Herbert. A. Fertig, Liang Fu, Bertrand Halperin, Udit Khanna, Ganpathy Murthy, Zlatko Papic, Edward H. Rezayi, and Efrat Shimshoni for helpful discussions. We thank Lu Li and Kuan-Wen Chen for providing guidance and a capacitance thermometer used to calibrate the temperature readings in cell 9 (31 T, 0.3 K) of the National High Magnetic Field Laboratory. **Funding**: The experiment is supported by the Department of Energy through grants DE-SC0022947 and by the National Science Foundation through NSF-DMR-1904986. The National Science Foundation grant supported the device fabrication. The Department of Energy grant supported the measurements, data analysis and manuscript writing. A. C. B acknowledges the financial support from the Science and Engineering Research Board (SERB) of the Department of Science and Technology (DST) via the Mathematical Research Impact Centric Support (MATRICS) Grant No. MTR/2023/000002. Computational portions of this research work were conducted using the Nandadevi and Kamet supercomputers maintained and supported by the Institute of Mathematical Science's High-Performance Computing Center. Some numerical calculations were performed using the DiagHam package [DiagHam], for which we are grateful to its authors, https://www.nick-ux.org/diagham. J. K. J. were supported in part by the U. S. Department of Energy, Office of Basic Energy Sciences, under Grant No. DE-SC0005042. K.W. and T.T. acknowledge support from the JSPS KAKENHI (Grant Numbers 21H05233 and 23H02052), CREST(JPMJCR24A5), JST, and World Premier International Research Center Initiative (WPI), MEXT, Japan. Work performed at the National High Magnetic Field Laboratory was supported by the NSF through NSF-DMR-1644779 and the State of Florida.

The authors declare no competing interests.

The data that support the plots in the main text are available from Harvard Dataverse (https://doi.org/doi:10.7910/DVN/XXCJWU) and other findings of this study are available from the corresponding author upon reasonable request.

## APPENDIX A: DEVICE FABRICATION AND MEASUREMENT SETUPS

Devices 002, 011, and 015 used in this work were also used in Ref. [39], where detailed fabrication and characterization are given. Briefly, we use polypropylene carbonate (PPC) stamps to pick up thin h-BN/BLG/h-BN/graphite sequentially. After annealing the stack in $O_2$/Ar mixture at 450 ℃, we transfer another layer of graphite flake exfoliated on the PPC stamp onto the top of the stack. The Hall bar structure is patterned by e-beam lithography and reactive ion etching. The edge contact is made by the two-step etching protocol described in Ref. [39] and deposited with Cr/Au.

Electrical transport measurements were done using standard low-frequency lock-in techniques. All measurements in Fig. 3(c) and open circles in Fig. 8(e), Appendix C were taken in cell 9 of the National High Magnetic Field Laboratory (NHMFL) using a He-3 cryostat. Solid circles in Fig. 8(e), Appendix C were taken in SCM-1 of the NHMFL. A small discrepancy between the two sets of data is attributed to the lower electron temperature in SCM-1. Blue squares (device 015) in Fig. 9, Appendix C were taken in SCM-4 (dilution fridge, up to 28 T) of the NHMFL.

## APPENDIX B: INTERACTIONS USED IN THE EXACT DIAGONALIZATION STUDIES

For the isospin that combines the valley and orbital indices ($|\mp 1\rangle$, $|\pm 0\rangle$), the Coulomb interaction is not SU(2) symmetric because it depends on the orbital degree of freedom. It can be fully characterized by the Haldane pseudopotentials [49], which are the energies of two electrons in a definite relative angular momentum $m$. The general form in the planar disk geometry is given by (the magnetic length $l_B$ is set to unity)

$$V_m^{\mathrm{N,N'}}(\theta) = \int dq\, F^{\mathrm{N,N'}}(q) e^{-q^2} L_m(q^2) \qquad \text{(Eq. B1)}$$

where q is the wave number, $L_k(x)$ is the kth Laguerre polynomial, and $F^{\mathrm{N,N'}}(q)$ is called the form-factor with N and N' being the orbital indices of the two electrons. The form factors are given by

$$F^{0,0}(q) = 1$$

$$F^{1,1}(\theta, \mathrm{q}) = \left[\sin^2(\theta)\, \mathrm{L}_1\left(\frac{\mathrm{q}^2}{2}\right) + \cos^2(\theta)\, \mathrm{L}_0\left(\frac{\mathrm{q}^2}{2}\right)\right]^2 \qquad \text{(Eq. B2)}$$

$$F^{0,1}(\theta, q) = L_0\left(\frac{q^2}{2}\right)\left[sin^2(\theta)\, L_1\left(\frac{q^2}{2}\right) + cos^2(\theta)\, L_0\left(\frac{q^2}{2}\right)\right]$$

where the $B$-dependent parameter $\theta$ can be estimated as $tan(\theta) \approx t l_B/\sqrt{2}\hbar v_F$, where $t$ is the hopping integral and $v_F$ is the Fermi velocity. For graphene, the typical Fermi velocity is $v_F$ =$10^6$ m/s, and taking $t$ = 350 meV, as obtained from DFT calculations at zero field [50], the magnetic field $B$ and $\theta$ are related through $B$ = 93.06$[\cot(\theta)]^2$ [T]. In Eq. (B2) it suffices to consider the range $0 \leq \theta \leq \pi/2$ since the form-factor only depends on $\sin^2(\theta)$. For $\theta$ = 0, or equivalently $B \longrightarrow \infty$, we recover the form factor for the n = 0 LL of GaAs.

The three sets of pseudopotentials are given by:

$$V_m^{0,0} = \frac{\Gamma\left(m + \frac{1}{2}\right)}{2\Gamma(m+1)}$$

$$V_m^{1,1}(\theta) = \left[\sin^2(\theta)\, L_1\left(\frac{\mathrm{q}^2}{2}\right) + \cos^2(\theta)\, L_0\left(\frac{\mathrm{q}^2}{2}\right)\right]^2 \qquad \text{(Eq. B3)}$$

$$V_m^{0,1}(\theta) = \frac{\sqrt{\pi}}{32}\left[16\, {}_2F_1\left(\frac{1}{2}, -m; 1; 1\right) - 8\, {}_2F_1\left(\frac{3}{2}, -m; 1; 1\right)\sin^2(\theta) + 3\, {}_2F_1\left(\frac{5}{2}, -m; 1; 1\right)\sin^4(\theta)\right]$$

where $\Gamma(x)$ is the Gamma function and ${}_2F_1$ is the Gauss hypergeometric function. The analogous spherical pseudopotentials can be obtained by following the derivation outlined in Refs. [51-54]. These pseudopotentials are used in the exact diagonalization computations of the ground state.

## APPENDIX C: ADDITIONAL FIGURES

Figure 5: The appearance of the $D^*$ states at different filling factors, magnetic fields and in different devices

Figure 6: The complete magnetic field evolution data of the $\nu = 7/5$ $D^*$ state in device 002

Figure 7: The magnetic field evolution of the $D^*$state at ν = 7/5, 10/7, and 13/9

Figure 8: Extracting the energy gaps of the ν = 7/5 $D^*$ state

Figure 9: Extracting the energy gaps of the N = 1 phase of 7/5

Figure 10: Extracting the energy gaps of the N = 0 phase of 7/5 and composite fermion mass

Figure 11: Theoretical calculations of the anisotropic two-component states at ν = 3/7 and 3/5

Figure 12: A comparison of the magnetic field-dependent thermodynamic energies of the various candidate states at filling factor 2/5.

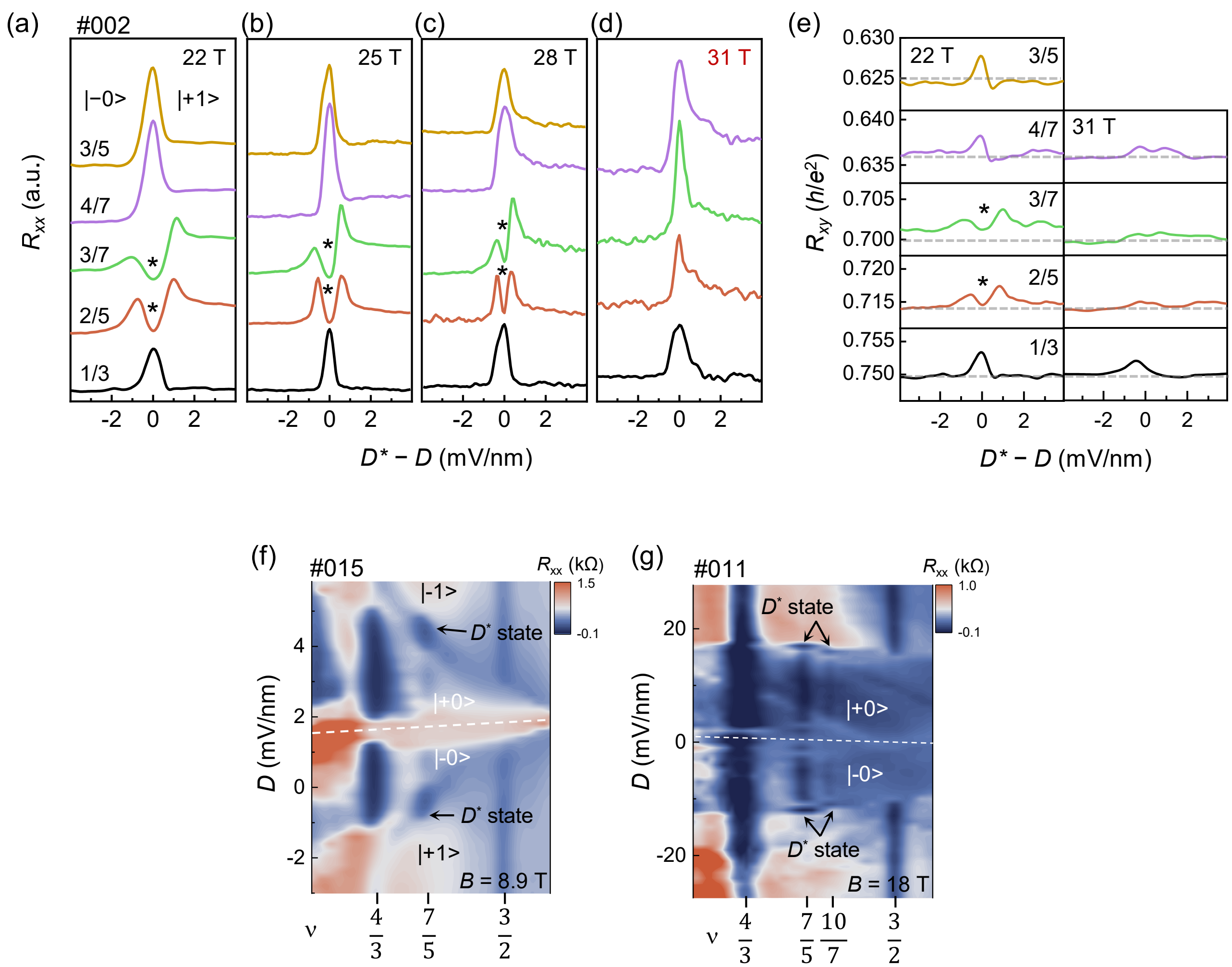

Fig.5. (a) – (d), $R_{xx}(D)$ traces taken along fixed filling factors and in different magnetic fields from 22 to 31 T as labeled. Partial fillings $\tilde{\nu} = \nu - 1 = 1/3$ (black), 2/5 (red), 3/7(green), 4/7(violet), and 3/5(brown). Traces are vertically stacked for clarity. (e) The corresponding $R_{xy}(D)$ traces taken at 22 T and 31 T. The dashed lines indicate the expected Hall resistance value at each filling factor. All filling factors exhibit FQH states on the $|-0\rangle$ and $|+1\rangle$ LLs. The $D^*$ state (marked by the * symbol) forms at 2/5 and 3/7 but not at other filling factors. The $D^*$ state disappears completely at $B$ = 31 T. $T$ = 0.33 K. From device 002.

(f) and (g), False color map of $R_{xx}(D, \nu)$ in devices 015 and 011 respectively showing the presence of the $D^*$ states at ν = 7/5 and 10/7. The white dashed line traces the true $D = 0$ locations in each device. $B = 8.9$ T and $T = 0.33$ K in (f). $B = 18$ T and $T = 20$ mK in (g).

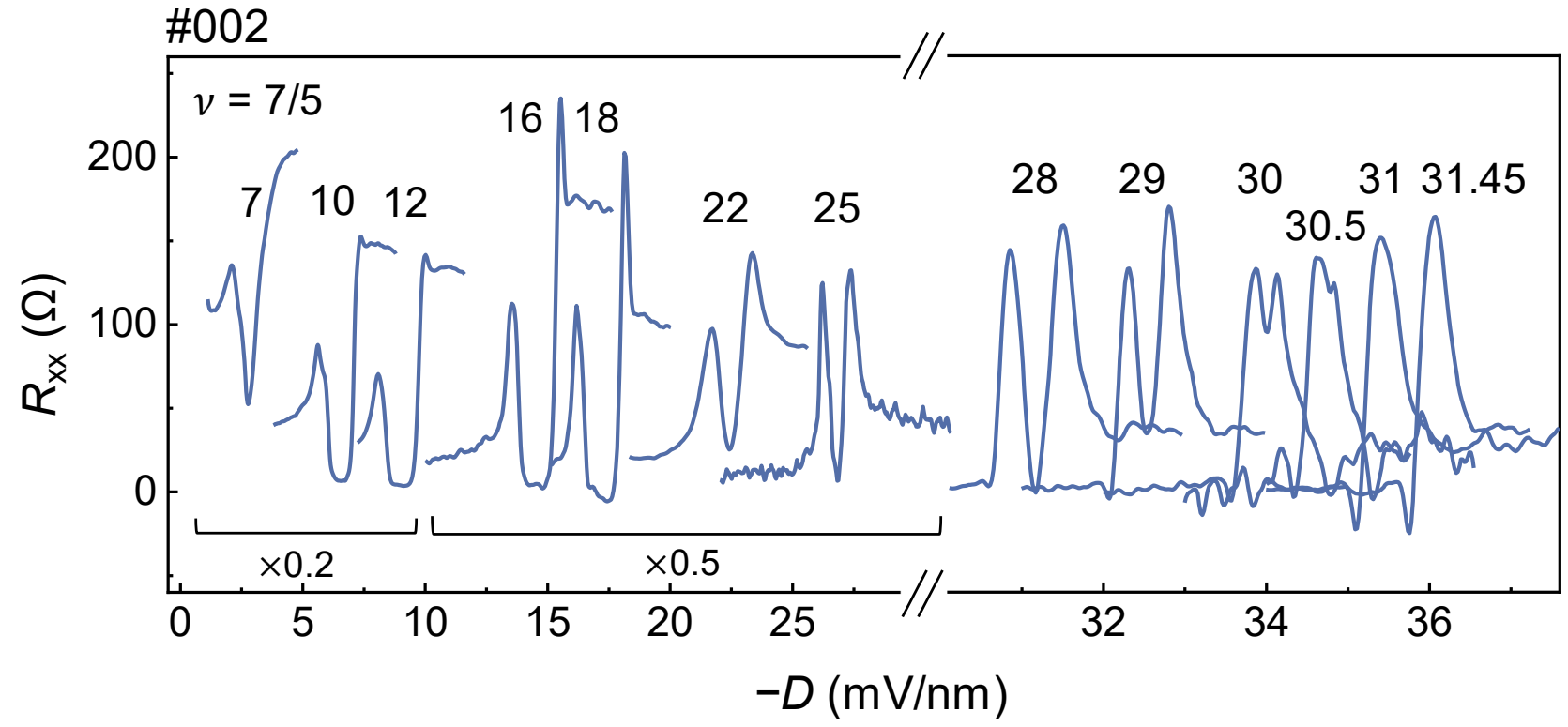

Fig. 6. $R_{xx}^{7/5}(D)$ sweeps at fixed $B$-fields as labeled in the plot. From device 002.

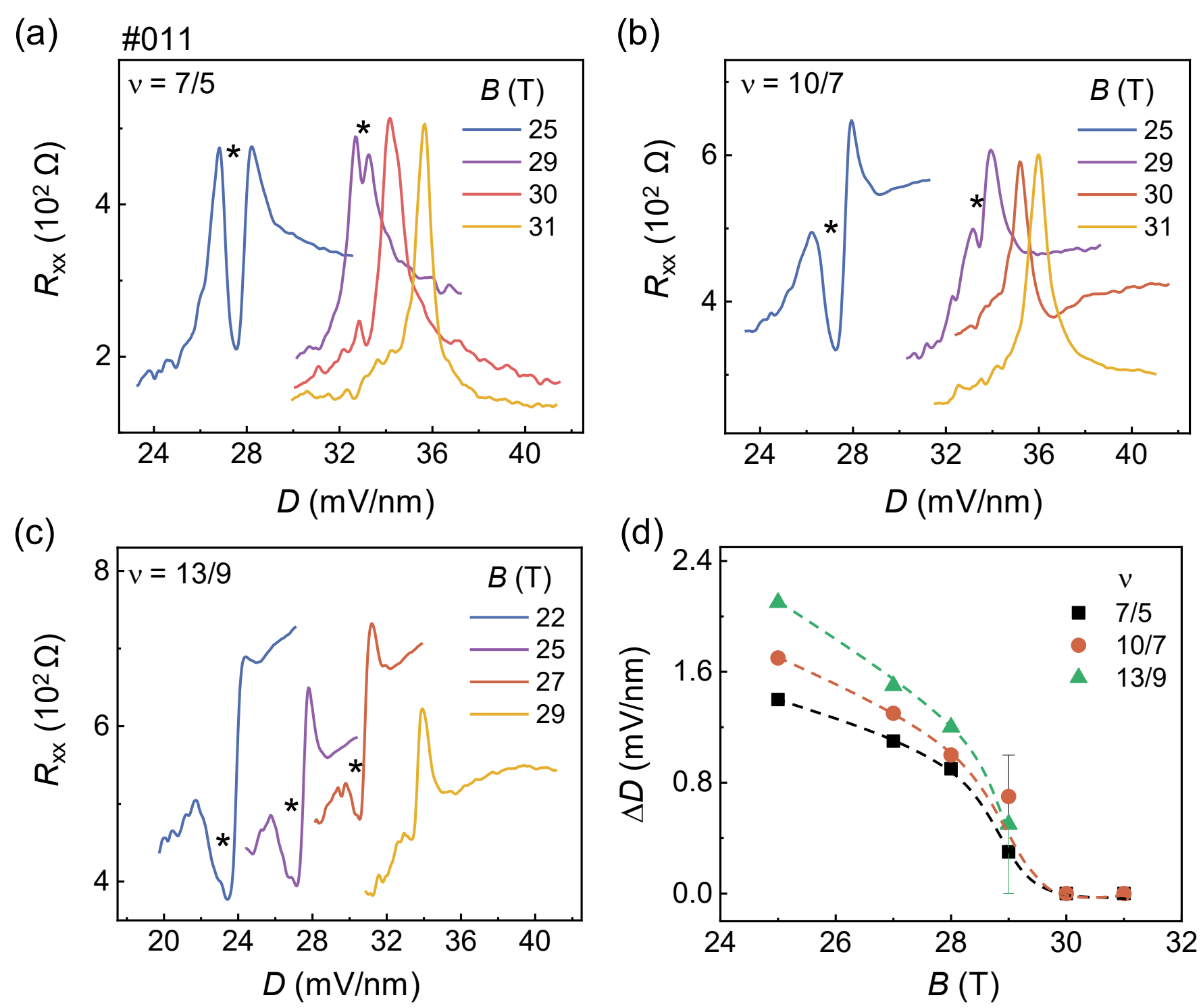

Fig. 7. Magnetic field evolution of the $D^*$ state at ν = 7/5, 10/7, 13/9 in device 011. (a) – (c), $R_{xx}^{7/5}(D)$, $R_{xx}^{10/7}(D)$ and $R_{xx}^{13/9}(D)$ sweeps at selected $B$-fields as labeled in the plot. The $D^*$ state is marked by a * symbol. $\Delta D$ is as defined in the main Fig. 2. (d) $\Delta D(B)$ for ν = 7/5 (square), 10/7 (circle), and 13/9 (triangle). All three vanish together near 30 T. The dashed lines are guides to the eye.

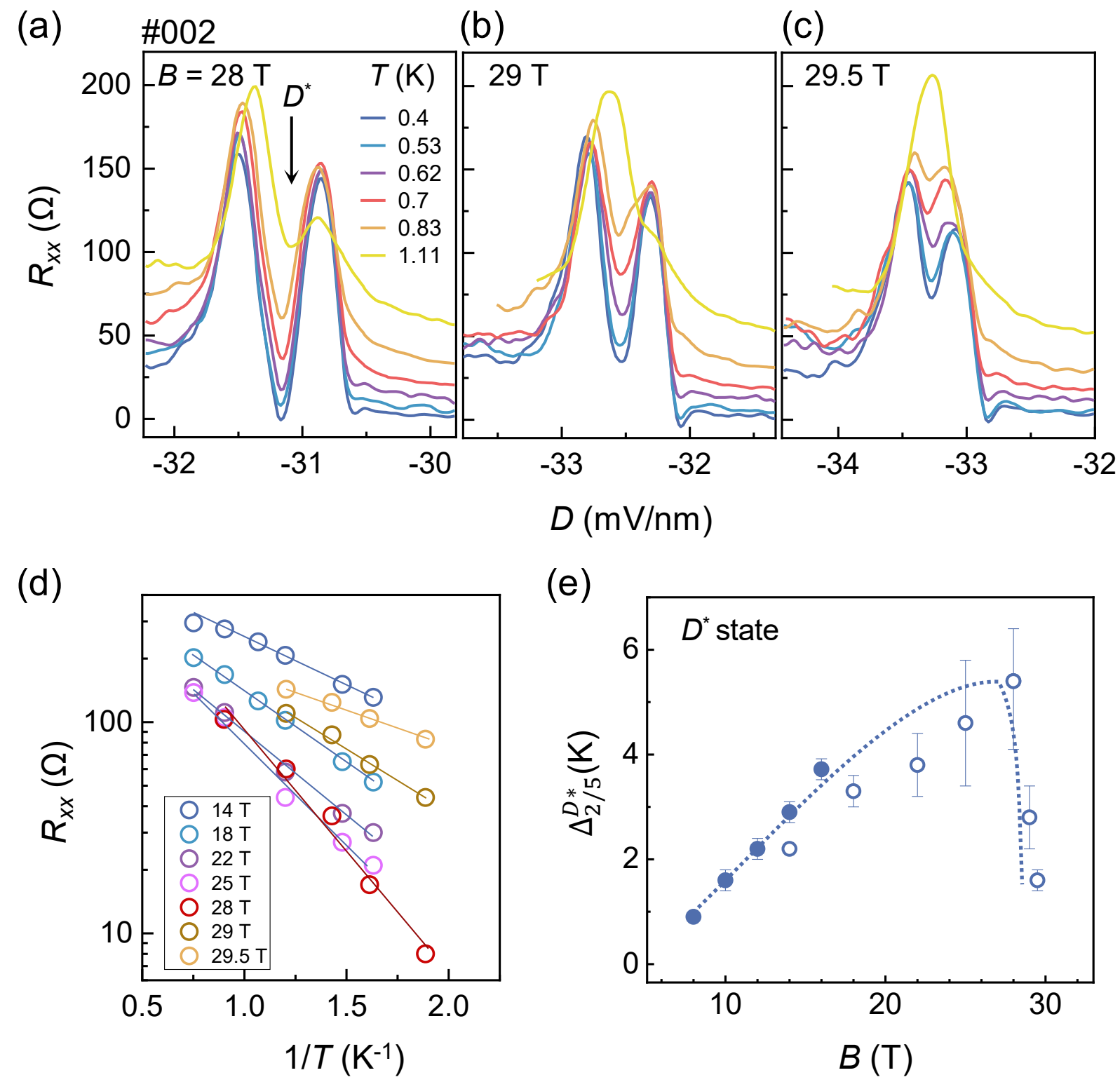

Fig. 8. Energy gap of the $D^*$ phase of 7/5. (a) – (c), $T$-dependent $R_{xx}^{7/5}(D)$ traces taken at $B$ = 28, 29, and 29.5 T respectively. As the phase space of the $D^*$ state narrows, it is also gradually destabilized likely due to an increasing mixture of N = 0 and N = 1 domains. (d) Arrhenius plots and fits that produced the gray open squares shown in Fig. 3(c). e, $\Delta_{2/5}^{D^*}$ as a function of $B$ including results obtained from (d) (open circle) and additional measurements taken in a different cryostat (solid circle). A small systematic difference between the two sets is due to the different temperature readout schemes. From device 002.

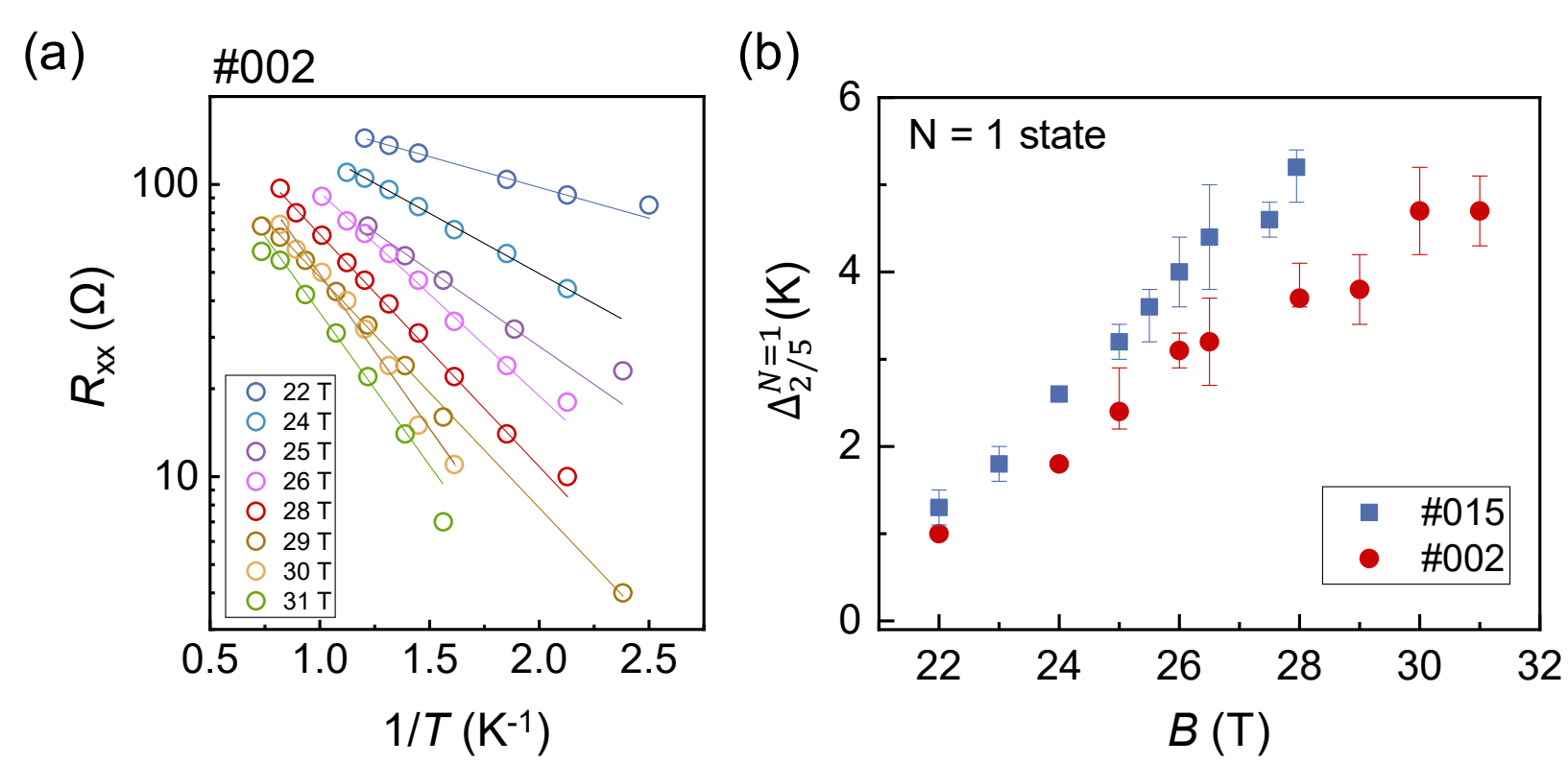

Fig. 9. Energy gap of the N = 1 phase of 7/5. (a) Arrhenius plots and fits for the N = 1 state at different $B$-fields as labeled. From device 002. (b) $\Delta_{2/5}^{N=1}$ as a function of $B$ from the fits in (a) (circle) and from device 015 (square) using similar measurements. Both devices show that $\Delta_{2/5}^{N=1}$ is very small at low and intermediate $B$-fields but increases rapidly with $B$ once the field reaches 20 T.

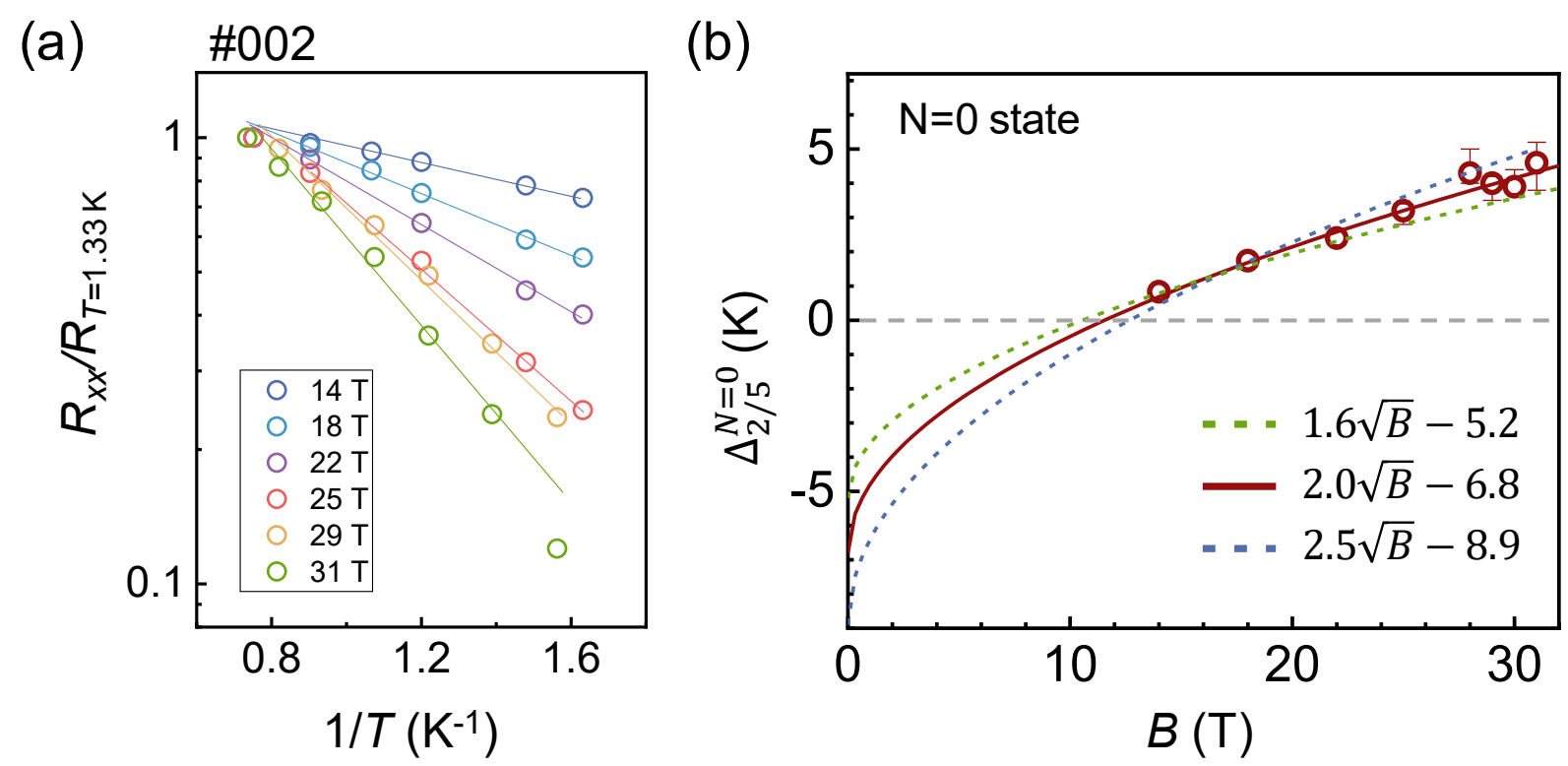

Fig. 10. Energy gap of the N = 0 phase of 7/5. (a) Arrhenius plots and fits for the N = 0 state at different *B*-fields as labeled. We normalized $R_{xx}(T)$ by $R_{xx}$ measured at $T = 1.33$ K to better compare the data taken at different magnetic fields. (b) $\Delta_{2/5}^{N=0}$ as a function of $B$ extracted from a. The best fit to data is given by the red solid line, which corresponds to $\Delta_{2/5}^{N=0} = 2.0\sqrt{B} - \Gamma$, where $\Gamma = 6.8$ K is the disorder broadening energy scale consistent with our previous assessment of the bulk disorder level in this device [39]. Also shown are the high and low bounds of possible fits. They correspond to $2.5\sqrt{B} - 8.9$ (blue dashed line) and $1.6\sqrt{B} - 5.2$ (green dashed line) respectively. Using the CF model, we write $\Delta_{7/5} = \hbar e B_{eff}/m_a^* - \Gamma$, where $B_{eff} = 3(B_{7/5} - B_{3/2}) = B_{7/5}/5$ is the effective magnetic field at ν = 7/5 and $m_a^* = \alpha m_e\sqrt{B}$ is the effective activation CF mass. The fits correspond to $\alpha = 0.13 \pm 0.03$.

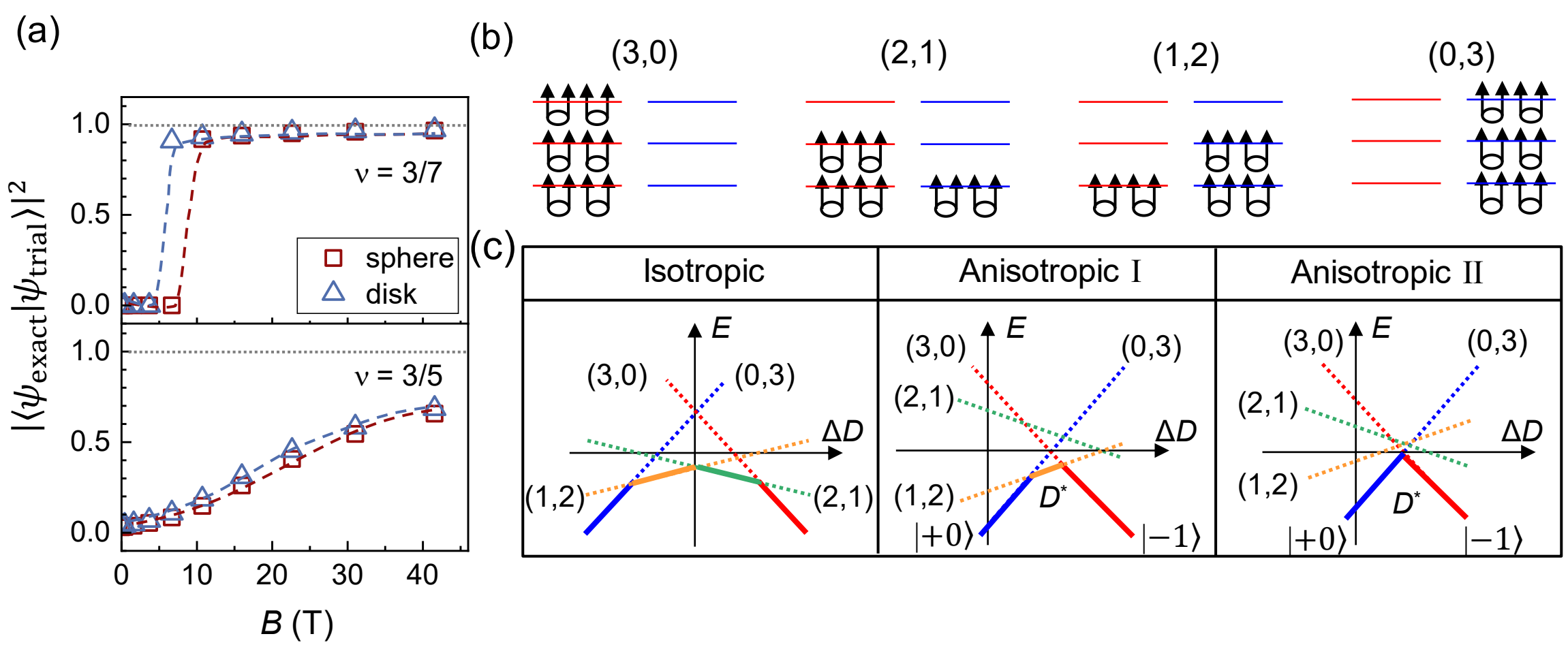

Fig. 11. Theoretical calculations of anisotropic two-component states at ν = 3/7 and 3/5. (a) Overlaps of the partially-polarized Jain state ($|\psi_{trial}\rangle$) with the exact Coulomb ground state ($|\psi_{exact}\rangle$) in the $E = 0$ LL of BLG as a function of the magnetic field for $\nu = 3/7$ (upper panel) and 3/5 (lower panel). Calculations are performed for n electrons in the spherical geometry using the spherical (red sphere) and disk (blue triangle) pseudopotentials. The electron number n = 11 for ν = 3/7 and 14 for ν = 3/5. (b) CF Λ level filling diagram for the 3/7 state showing four potential isospin configurations. The $|-1\rangle$ state corresponds to (3, 0). The $|+0\rangle$ state corresponds to (0, 3). The (2, 1) and (1, 2) are partially isospin polarized states. (c) Schematic energy diagrams showing the evolution of the four configurations in a *D*-field and the resulting

ground states. In an isotropic interaction, both the (2, 1) and the (1, 2) manifest. This is indeed observed in Ref. [39], where the two-component spinor is $(|+0\rangle, |-0\rangle)$. Similar to the case of 2/5, anisotropic interactions change the energies of the four states. The middle diagram depicts a scenario where only the (1, 2) state manifests. This is likely the $D^*$ state observed in experiment. The last diagram shows how the system may transition from $|+0\rangle$ to $|-1\rangle$ directly. This may correspond to the disappearance of the 3/7 $D^*$ state at large $B$.

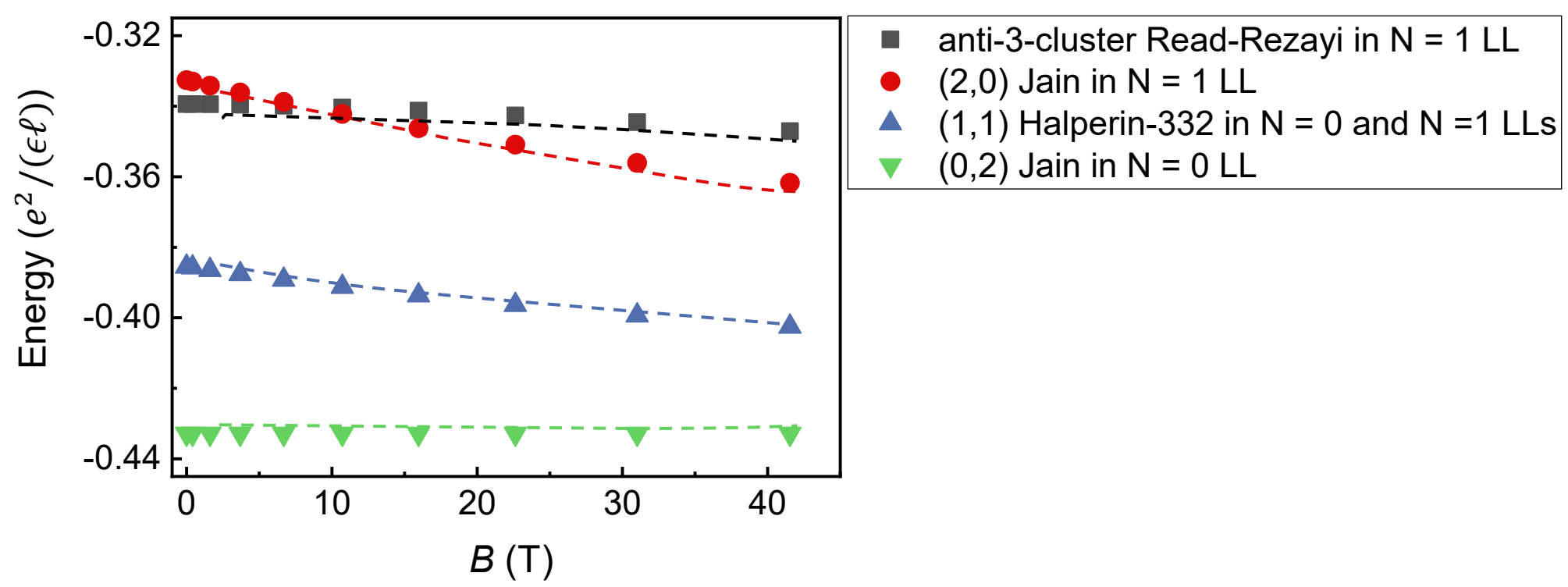

Fig. 12. Magnetic field-dependent thermodynamic energies of the various candidate states at filling factor 2/5. The calculations are performed for the zeroth LL of bilayer graphene in the spherical geometry. The two isospin components correspond to $|-1\rangle$ and $|+0\rangle$.

## Reference

[1] J. K. Jain, *Composite Fermions* (Cambridge University Press, Cambridge, 2007). https://doi.org/10.1017/CBO9780511607561

[2] B. Halperin and J. K. Jain, *Fractional Quantum Hall Effects: New Developments* (World Scientific, 2020). https://doi.org/10.1142/11751

[3] R. L. Willett, The quantum Hall effect at 5/2 filling factor, Reports on Progress in Physics **76**, 076501 (2013). https://doi.org/10.1088/0034-4885/76/7/076501

[4] J. S. Xia, W. Pan, C. L. Vicente, E. D. Adams, N. S. Sullivan, H. L. Stormer, D. C. Tsui, L. N. Pfeiffer, K. W. Baldwin, and K. W. West, Electron correlation in the second Landau level: A competition between many nearly degenerate quantum phases, Physical Review Letters **93**, 176809 (2004). https://doi.org/10.1103/PhysRevLett.93.176809

[5] A. Kumar, G. A. Csathy, M. J. Manfra, L. N. Pfeiffer, and K. W. West, Non-conventional Odd-Denominator Fractional Quantum Hall States in the Second Landau Level, Physical Review Letters **105**, 246808, 246808 (2010). https://doi.org/10.1103/PhysRevLett.105.246808

[6] N. Read and E. Rezayi, Beyond paired quantum Hall states: Parafermions and incompressible states in the first excited Landau level, Physical Review B **59**, 8084 (1999). https://doi.org/10.1103/PhysRevB.59.8084

[7] E. Grosfeld and K. Schoutens, Non-Abelian Anyons: When Ising Meets Fibonacci, Physical Review Letters **103**, 076803 (2009). https://doi.org/10.1103/PhysRevLett.103.076803

[8] L. Hormozi, N. E. Bonesteel, and S. H. Simon, Topological Quantum Computing with Read-Rezayi States, Physical Review Letters **103**, 160501 (2009). https://doi.org/10.1103/PhysRevLett.103.160501

[9] W. Zhu, S. S. Gong, F. D. M. Haldane, and D. N. Sheng, Fractional Quantum Hall States at ν = 13/5 and 12/5 and Their Non-Abelian Nature, Physical Review Letters **115**, 126805 (2015). https://doi.org/10.1103/PhysRevLett.115.126805
[10] J. A. Hutasoit, A. C. Balram, S. Mukherjee, Y. H. Wu, S. S. Mandal, A. Wójs, V. Cheianov, and J. K. Jain, The enigma of the ν = 2+3/ 8 fractional quantum Hall effect, Physical Review B **95** (2017). https://doi.org/10.1103/PhysRevB.95.125302
[11] W. N. Faugno, J. K. Jain, and A. C. Balram, Non-Abelian fractional quantum Hall state at 3/7-filled Landau level, Physical Review Research **2**, 033223 (2020). https://doi.org/10.1103/PhysRevResearch.2.033223
[12] A. C. Balram, Transitions from Abelian composite fermion to non-Abelian parton fractional quantum Hall states in the zeroth Landau level of bilayer graphene, Physical Review B **105**, L121406 (2022). https://doi.org/10.1103/PhysRevB.105.L121406
[13] Z. Papić, F. D. M. Haldane, and E. H. Rezayi, Quantum Phase Transitions and the ν = 5/2 Fractional Hall State in Wide Quantum Wells, Physical Review Letters **109**, 266806 (2012). https://doi.org/10.1103/PhysRevLett.109.266806
[14] M. Barkeshli, C. Nayak, Z. Papic, A. Young, and M. Zaletel, Topological Exciton Fermi Surfaces in Two-Component Fractional Quantized Hall Insulators, Physical Review Letters **121** (2018). https://doi.org/10.1103/PhysRevLett.121.026603
[15] T. Jolicoeur, C. Toke, and I. Sodemann, Quantum Hall ferroelectric helix in bilayer graphene, Physical Review B **99** (2019). https://doi.org/10.1103/PhysRevB.99.115139
[16] Z. Zhu, D. N. Sheng, and I. Sodemann, Widely Tunable Quantum Phase Transition from Moore-Read to Composite Fermi Liquid in Bilayer Graphene, Physical Review Letters **124**, 097604 (2020). https://doi.org/10.1103/PhysRevLett.124.097604
[17] U. Khanna, K. Huang, G. Murthy, H. A. Fertig, K. Watanabe, T. Taniguchi, J. Zhu, and E. Shimshoni, Phase diagram of the ν = 2 quantum Hall state in bilayer graphene, Physical Review B **108** (2023). https://doi.org/10.1103/PhysRevB.108.L041107
[18] Y. Liu, D. Kamburov, M. Shayegan, L. N. Pfeiffer, K. W. West, and K. W. Baldwin, Anomalous Robustness of the ν = 5/2 Fractional Quantum Hall State near a Sharp Phase Boundary, Physical Review Letters **107**, 176805, 176805 (2011). https://doi.org/10.1103/PhysRevLett.107.176805
[19] J. Falson, D. Tabrea, D. Zhang, I. Sodemann, Y. Kozuka, A. Tsukazaki, M. Kawasaki, K. von Klitzing, and J. H. Smet, A cascade of phase transitions in an orbitally mixed half-filled Landau level, Science Advances **4** (2018). https://doi.org/10.1126/sciadv.aat8742
[20] R. R. Du, A. S. Yeh, H. L. Stormer, D. C. Tsui, L. N. Pfeiffer, and K. W. West, Fractional Quantum Hall Effect around ν = 3/2: Composite Fermions with a Spin, Physical Review Letters **75**, 3926 (1995). https://doi.org/10.1103/PhysRevLett.75.3926
[21] B. E. Feldman, A. J. Levin, B. Krauss, D. A. Abanin, B. I. Halperin, J. H. Smet, and A. Yacoby, Fractional Quantum Hall Phase Transitions and Four-Flux States in Graphene, Physical Review Letters **111** (2013). https://doi.org/10.1103/PhysRevLett.111.076802
[22] Y. Zeng, J. I. A. Li, S. A. Dietrich, O. M. Ghosh, K. Watanabe, T. Taniguchi, J. Hone, and C. R. Dean, High-Quality Magnetotransport in Graphene Using the Edge-Free Corbino Geometry, Physical Review Letters **122**, 137701 (2019). https://doi.org/10.1103/PhysRevLett.122.137701
[23] K. Huang, P. Wang, L. N. Pfeiffer, K. W. West, K. W. Baldwin, Y. Liu, and X. Lin, Resymmetrizing Broken Symmetry with Hydraulic Pressure, Physical Review Letters **123**, 206602 (2019). https://doi.org/10.1103/PhysRevLett.123.206602
[24] X. G. Wu, G. Dev, and J. K. Jain, Mixed-Spin Incompressible States in the Fractional Quantum Hall-Effect, Physical Review Letters **71**, 153 (1993). https://doi.org/10.1103/PhysRevLett.71.153

[25] I. V. Kukushkin, K. v. Klitzing, and K. Eberl, Spin Polarization of Composite Fermions: Measurements of the Fermi Energy, Physical Review Letters **82**, 3665 (1999). https://doi.org/10.1103/PhysRevLett.82.3665
[26] S. C. Davenport and S. H. Simon, Spinful composite fermions in a negative effective field, Physical Review B **85**, 245303 (2012). https://doi.org/10.1103/PhysRevB.85.245303
[27] A. C. Balram, C. Tőke, A. Wójs, and J. K. Jain, Fractional quantum Hall effect in graphene: Quantitative comparison between theory and experiment, Physical Review B **92**, 075410 (2015). https://doi.org/10.1103/PhysRevB.92.075410
[28] J. P. Eisenstein, G. S. Boebinger, L. N. Pfeiffer, K. W. West, and S. He, New Fractional Quantum Hall State in Double-layer Two-Dimensional Electron Systems, Physical Review Letters **68**, 1383 (1992). https://doi.org/10.1103/PhysRevLett.68.1383
[29] Y. W. Suen, L. W. Engel, M. B. Santos, M. Shayegan, and D. C. Tsui, Observation of a $\nu = 1/2$ Fractional Quantum Hall State in a Double-layer Electron System, Physical Review Letters **68**, 1379 (1992). https://doi.org/10.1103/PhysRevLett.68.1379
[30] J. I. A. Li, T. Taniguchi, K. Watanabe, J. Hone, and C. R. Dean, Excitonic superfluid phase in double bilayer graphene, Nature Physics **13**, 751 (2017). https://doi.org/10.1038/nphys4140
[31] X. Liu, J. I. A. Li, K. Watanabe, T. Taniguchi, J. Hone, B. I. Halperin, P. Kim, and C. R. Dean, Crossover between strongly coupled and weakly coupled exciton superfluids, Science **375**, 205 (2022). https://doi.org/10.1126/science.abg1110
[32] Q. H. Shi, E. M. Shih, D. Rhodes, B. Kim, K. Barmak, K. Watanabe, T. Taniguchi, Z. Papic, D. A. Abanin, J. Hone, and C. R. Dean, Bilayer $WSe_2$ as a natural platform for interlayer exciton condensates in the strong coupling limit, Nature Nanotechnology **17**, 577 (2022). https://doi.org/10.1038/s41565-022-01104-5
[33] N. C. Bishop, M. Padmanabhan, K. Vakili, Y. P. Shkolnikov, E. P. De Poortere, and M. Shayegan, Valley polarization and susceptibility of composite fermions around a filling factor $\nu = 3/2$, Physical Review Letters **98** (2007). https://doi.org/10.1103/PhysRevLett.98.266404
[34] B. M. Hunt, J. I. A. Li, A. A. Zibrov, L. Wang, T. Taniguchi, K. Watanabe, J. Hone, C. R. Dean, M. Zaletel, R. C. Ashoori, and A. F. Young, Direct measurement of discrete valley and orbital quantum numbers in bilayer graphene, Nature Communications **8**, 948, 948 (2017). https://doi.org/10.1038/s41467-017-00824-w
[35] J. I. A. Li, C. Tan, S. Chen, Y. Zeng, T. Taniguehi, K. Watanabe, J. Hone, and C. R. Dean, Even-denominator fractional quantum Hall states in bilayer graphene, Science **358**, 648 (2017). https://doi.org/10.1126/science.aao2521
[36] A. A. Zibrov, C. Kometter, H. Zhou, E. M. Spanton, T. Taniguchi, K. Watanabe, M. P. Zaletel, and A. F. Young, Tunable interacting composite fermion phases in a half-filled bilayer-graphene Landau level, Nature **549**, 360 (2017). https://doi.org/10.1038/nature23893
[37] J. Li, Y. Tupikov, K. Watanabe, T. Taniguchi, and J. Zhu, Effective Landau Level Diagram of Bilayer Graphene, Physical Review Letters **120**, 047701 (2018). https://doi.org/10.1103/PhysRevLett.120.047701
[38] A. A. Zibrov, E. M. Spanton, H. Zhou, C. Kometter, T. Taniguchi, K. Watanabe, and A. F. Young, Even-denominator fractional quantum Hall states at an isospin transition in monolayer graphene, Nature Physics **14**, 930 (2018). https://doi.org/10.1038/s41567-018-0190-0
[39] K. Huang, H. L. Fu, D. R. Hickey, N. Alem, X. Lin, K. Watanabe, T. Taniguchi, and J. Zhu, Valley Isospin Controlled Fractional Quantum Hall States in Bilayer Graphene, Physical Review X **12** (2022). https://doi.org/10.1103/PhysRevX.12.049901

[40] Y. Liu, S. Hasdemir, D. Kamburov, A. L. Graninger, M. Shayegan, L. N. Pfeiffer, K. W. West, K. W. Baldwin, and R. Winkler, Even-denominator fractional quantum Hall effect at a Landau level crossing, Physical Review B **89** (2014). https://doi.org/10.1103/PhysRevB.89.165313
[41] C. Y. Wang, A. Gupta, Y. J. Chung, L. N. Pfeiffer, K. W. West, K. W. Baldwin, R. Winkler, and M. Shayegan, Highly Anisotropic Even-Denominator Fractional Quantum Hall State in an Orbitally Coupled Half-Filled Landau Level, Physical Review Letters **131** (2023). https://doi.org/10.1103/PhysRevLett.131.056302
[42] B. I. Halperin, Theory of the Quantized Hall Conductance, Helvetica Physica Acta **56**, 75 (1983). https://doi.org/10.5169/seals-115362
[43] B. I. Halperin, P. A. Lee, and N. Read, Theory of the Half-Filled Landau-Level, Physical Review B **47**, 7312 (1993). https://doi.org/10.1103/PhysRevB.47.7312
[44] K. Park and J. K. Jain, Phase Diagram of the Spin Polarization of Composite Fermions and a New Effective Mass, Physical Review Letters **80**, 4237 (1998). https://doi.org/10.1103/PhysRevLett.80.4237
[45] X. Liu, G. Farahi, C.-L. Chiu, Z. Papic, K. Watanabe, T. Taniguchi, M. P. Zaletel, and A. Yazdani, Visualizing broken symmetry and topological defects in a quantum Hall ferromagnet, Science **375**, 321 (2022). https://doi.org/10.1126/science.abm3770
[46] J. Li, H. Fu, Z. Yin, K. Watanabe, T. Taniguchi, and J. Zhu, Metallic Phase and Temperature Dependence of the $\nu = 0$ Quantum Hall State in Bilayer Graphene, Physical Review Letters **122**, 097701 (2019). https://doi.org/10.1103/PhysRevLett.122.097701
[47] P. Maher, C. R. Dean, A. F. Young, T. Taniguchi, K. Watanabe, K. L. Shepard, J. Hone, and P. Kim, Evidence for a spin phase transition at charge neutrality in bilayer graphene, Nature Physics **9**, 154 (2013). https://doi.org/10.1038/nphys2528
[48] H. Fu, K. Huang, K. Watanabe, T. Taniguchi, and J. Zhu, Gapless Spin Wave Transport through a Quantum Canted Antiferromagnet, Physical Review X **11**, 021012 (2021). https://doi.org/10.1103/PhysRevX.11.021012
[49] F. D. M. Haldane, Fractional Quantization of the Hall-Effect - a Hierarchy of Incompressible Quantum Fluid States, Physical Review Letters **51**, 605 (1983). https://doi.org/10.1103/PhysRevLett.51.605
[50] J. Jung and A. H. MacDonald, Accurate tight-binding models for the $\pi$ bands of bilayer graphene, Physical Review B **89** (2014). https://doi.org/10.1103/PhysRevB.89.035405
[51] A. C. Balram, C. Tőke, A. Wójs, and J. K. Jain, Spontaneous polarization of composite fermions in the $n = 1$ Landau level of graphene, Physical Review B **92**, 205120 (2015). https://doi.org/10.1103/PhysRevB.92.205120
[52] M. Arciniaga and M. R. Peterson, Landau level quantization for massless Dirac fermions in the spherical geometry: Graphene fractional quantum Hall effect on the Haldane sphere, Physical Review B **94** (2016). https://doi.org/10.1103/PhysRevB.94.035105
[53] W. H. Hsiao, Landau quantization of multilayer graphene on a Haldane sphere, Physical Review B **101** (2020). https://doi.org/10.1103/PhysRevB.101.155310
[54] R. K. Dora and A. C. Balram, Competition between fractional quantum Hall liquid and electron solid phases in the Landau levels of multilayer graphene, Physical Review B **108** (2023). https://doi.org/10.1103/PhysRevB.108.235153